\documentclass[journal,comsoc]{IEEEtran}
\usepackage{flushend}
\usepackage{cite}
\usepackage{booktabs}
\usepackage{graphicx}
\usepackage{float}
\usepackage{subfigure}
\usepackage{color}
\usepackage{times}
\usepackage{amsmath,amssymb}
\usepackage[ruled,vlined]{algorithm2e}
\usepackage{multicol}
\usepackage{multirow}

\ifCLASSINFOpdf

\else

\fi

\usepackage{amsmath}

\usepackage[cmintegrals]{newtxmath}

\begin{document}

\title{The Robustness of Graph \emph{k}-shell Structure under Adversarial Attacks}

\author{Bo Zhou, 
        Yuqian Lv, Yongchao Mao, Jinhuan Wang, Shanqing Yu, and~Qi~Xuan,~\IEEEmembership{Member,~IEEE}
\thanks{This work was supported in part by the National Natural Science Foundation of China under Grants 61973273, by the Zhejiang Provincial Natural Science Foundation of China under Grants LR19F030001, and by the Research and Development Center of Transport Industry of New Generation of Artificial Intelligence Technology. \emph{(Corresponding authors: Qi Xuan.)}}    
\thanks{B. Zhou, Y. Lv, Y. Mao, J. Wang, S. Yu, and Q. Xuan are with the Institute of Cyberspace Security, College of Information Engineering, Zhejiang University of Technology, Hangzhou 310023, China (e-mail: xuanqi@zjut.edu.cn).}
\thanks{B. Zhou is also with Zhejiang Institute of Communications, Hangzhou 311112, China.}}


\maketitle

\begin{abstract}
The \emph{k}-shell decomposition plays an important role in unveiling the structural properties of a network, i.e., it is widely adopted to find the densest part of a network across a broad range of scientific fields, including Internet, biological networks, social networks etc. However, there arises concern about the robustness of the \emph{k}-shell structure when networks suffer from adversarial attacks. Here, we introduce and formalize the problem of \emph{k}-shell attack and develop an efficient strategy to attack the \emph{k}-shell structure by rewiring a small number of links. To the best of our knowledge, it is the first time to study the robustness of graph \emph{k}-shell structure under adversarial attacks. In particular, we propose a Simulated Annealing (SA) based \emph{k}-shell attack method and testify it on four real-world social networks. The extensive experiments validate that the \emph{k}-shell structure of a network is robust under random perturbation, but it is quite vulnerable under adversarial attack, e.g., in Dolphin and Throne networks, more than 40\% nodes change their \emph{k}-shell values when only 10\% links are changed based on our SA-based \emph{k}-shell attack. Such results suggest that a single structural feature could also be significantly disturbed when only a small fraction of links are changed purposefully in a network. Therefore, it could be an interesting topic to improve the robustness of various network properties against adversarial attack in the future.
\end{abstract}


\IEEEpeerreviewmaketitle

\section{Introduction}

\IEEEPARstart{N}{ETWORKS} or graphs can well represent various complex systems in our daily life~\cite{cook2006mining}, such as social networks~\cite{1995Social}, biological networks~\cite{junker2011analysis}, power networks~\cite{2004Optimization}, and financial networks~\cite{2014Complex}. Network science has been a very active field nowadays, due to its highly interdisciplinary attributes, and most of the tasks in this domain are related to network structure, such as link prediction~\cite{fu2018link}, node classification~\cite{dabhi2020nodenet}, network reconstruction~\cite{herrgaard2008consensus}. Considering the simplicity and broad applicability, \emph{k}-shell decomposition, 
which aims to find the largest subgraph of a network by recursively removing the nodes with degree less than \emph{k}, i.e., each node has \emph{k} neighbors in the remaining graph at least~\cite{2019k}, has become one of the most widely used methods to describe network structure. 
For instance, Altaf et al.~\cite{altaf2003prediction} proposed a procedure to predict the feature of functional-unknown proteins based upon \emph{k}-shell. Gaertler et al.~\cite{2004Dynamic} found \emph{k}-shell is useful in examining the Autonomous System (AS) Graph. Many researchers have proposed a number of methods~\cite{PhysRevE.70.056122,cha2010measuring,2010Thresholds,2011Modeling,al2017identification,wang2021sampling} based on \emph{k}-shell decomposition to identify the key nodes in a network or to measure the influence of users in online social networks. Miorandi et al.~\cite{Garas_2012} extended the \emph{k}-shell decomposition method to determine the impact of nodes on spreading epidemics in dynamic complex networks.

\begin{figure}[t]
	\centering
	\includegraphics[width=0.4\textwidth]{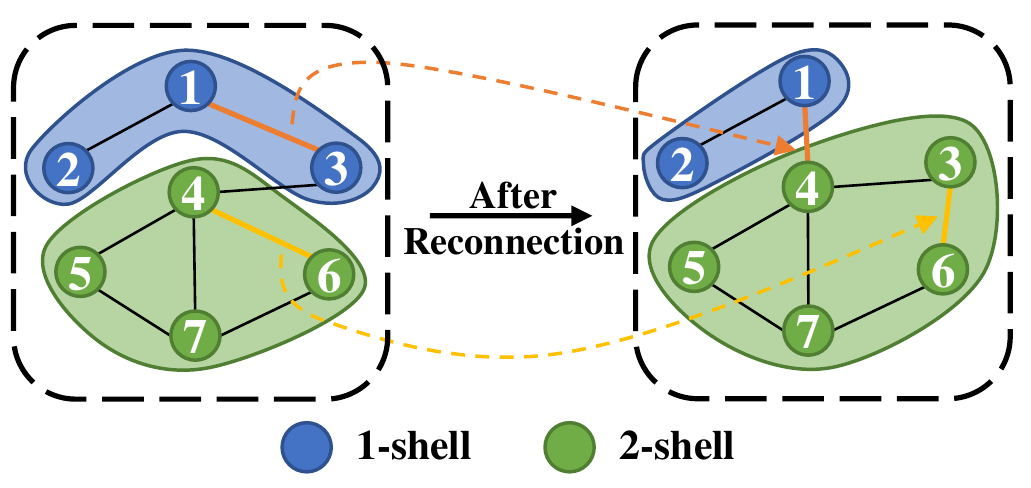}
	\caption{Node 3 is from 1-shell to 2-shell after rewiring two links, i.e., changing $e_{13}$ and $e_{46}$ to $e_{14}$ and $e_{36}$.}
	\label{figure_1}
	\label{fig:kshell}
\vspace{-5mm}
\end{figure}

With the rapid development of \emph{k}-shell applications in various fields, it would be an interesting question to ask: how robust the graph \emph{k}-shell structure is under various perturbations? 
Indeed, Kitsak et al.~\cite{kitsak2010identification} found that \emph{k}-shell decomposition is robust under random deletions of even up to 50\% of the links. However, there is still a lack of study on the robustness of the \emph{k}-shell structure under adversarial attack or purpose perturbation.
For example, as illustrated in Fig.~\ref{fig:kshell}, originally node 3 belongs to 1-shell; after rewiring two links, it turns to be 2-shell immediately. From a realistic point of view, in a credit scoring system, fraudsters can establish fake connections with several high-credit customers to evade the \emph{k}-shell decomposition to disturb the scoring system; and spammers can create fake followers easily to improve its centrality online, so as to increase the chance of fake news being 
spread to others. It is thus urgent to investigate the robustness of graph \emph{k}-shell structure under various adversarial attacks.

\begin{figure*}[!t]
	\centering
	\includegraphics[width=\textwidth]{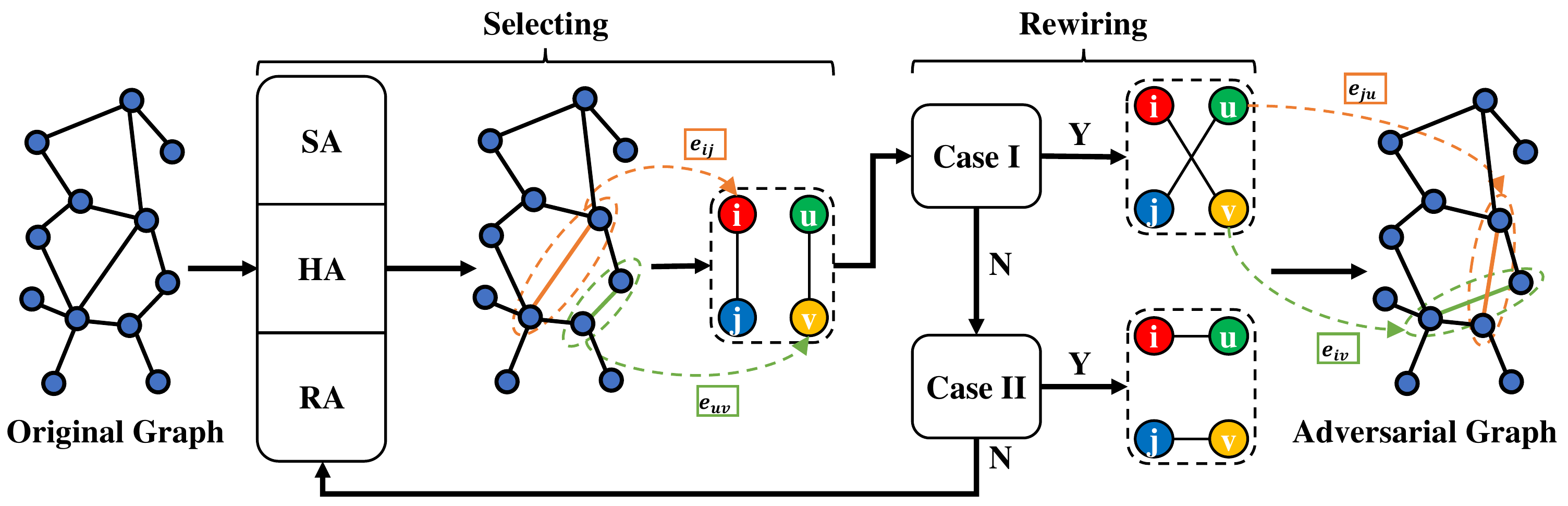}
	\caption{
	The overall framework of \emph{k}-shell attack. In selection module, two links $e_{ij}$ and $e_{uv}$ are selected. In rewiring module, if the two selected links satisfy \textbf{Case I} condition, we perform rewiring operation and the two links become $e_{iv}$ and $e_{ju}$, obtaining the adversarial graph. 
	}
	\label{figure1}
\vspace{-5mm}
\end{figure*}

Quite recently, numerous researchers devote themselves into the study of adversarial attacks on networks. Z\"{u}gner et al.~\cite{zugner2018adversarial} proposed a method to attack graph convolutional networks (GCN) for node classification. Yu et al.~\cite{8792200} proposed three heuristic strategies by rewiring to
attacked the RA link prediction method successfully. Chen et al.~\cite{chen2019ga} proposed a Genetic Algorithm (GA) based Q-Attack to destroy the network community structure. Xuan et al.~\cite{xuan2020adversarial} disturbed the degree distribution of networks to test the robustness of Broido and Clauset (BC) classification for scale-free networks in terms of statistical measures. Nonetheless, as far as we know, most of these studies focus on destroying machine learning algorithms on graphs but rarely concern the robustness of a particular feature, such as \emph{k}-shell, which is the basis of network analysis and graph machine algorithms in fact. We thus argue that it may be more crucial to investigate the robustness of particular structural features under adversarial attacks, which could provide deeper insight to understand the robustness of network structure from various aspects. 

In this paper, we propose the robustness of graph \emph{k}-shell structure under adversarial attacks for the first time, and make the following contributions:
\begin{enumerate}
\item We study the adversarial attack on \emph{k}-shell decomposition by rewiring links, for the first time.
\item We adopt the Simulated Annealing (SA) optimization method to select candidate links for rewiring, so as to realize the effective \emph{k}-shell attack.
\item We validate that the \emph{k}-shell structure of real-world networks are indeed much more vulnerable to our SA-based \emph{k}-shell attack, by comparing with Random rewiring Attack (RA) and Heuristic rewiring Attack (HA).
\end{enumerate}

The rest of paper is organized as follows. In Section~\ref{Sec:Method}, we introduce our methodology, including basic ideas, evaluation metrics, and our SA-based \emph{k}-shell attack. In Section~\ref{Sec:Experiment}, the experimental results on four real-world networks are presented to validate the effectiveness of our attack method. Finally, the paper is concluded in Section~\ref{Sec:Conclusion}.


\section{Methodology}\label{Sec:Method}
In this part, we will introduce our methods to testify the robustness of graph \emph{k}-shell structure. 

\subsection{Preliminaries and Ideas}\label{Sec:Def}
Given an undirected graph $G=(V,E)$ without self-loop, where $V$ is the set of nodes and $E\subseteq(V \times V)$ is the set of links, each node $v_i$ is assigned an index $k_i$ to represent its \emph{k}-shell value based on the \emph{k}-shell decomposition as a recursively pruning process. Nodes with the same value of $k$ constitute the \emph{k}-shell. And all the nodes with their indices not smaller than \emph{k} constitute \emph{k}-core. Each node has its own \emph{k}-core and \emph{k}-shell. The graph under adversarial attack is denoted as $G^{\prime}=(V^{\prime},E^{\prime})$, where $V^{\prime}=V$ and the \emph{k}-shell of node $v_i$ under adversarial attack is denoted as $k_i^{\prime}$. 

The whole attack process is illustrated in Fig. \ref{figure1}, where \textbf{Case I} and \textbf{Case II} are stated as below:
\begin{itemize}
    \item \textbf{Case I:} If link $e_{iv} \notin E$ and $e_{ju} \notin E$ , meanwhile, $v_i$ and $v_j$ are different from $v_v$ and $v_u$, respectively.
    \item \textbf{Case II:} If link $e_{iu} \notin E$ and $e_{jv} \notin E$, meanwhile, $v_i$ and $v_j$ are different from $v_u$ and $v_v$, respectively;

\end{itemize}

Then, the attack is realized by iteratively implementing the following two steps:
\begin{enumerate}
\item \textbf{Selecting:} We select a pair of links, i.e., $e_{ij}$ and $e_{uv}$, from the original graph through particular methods, including Random, Heuristic and SA-based \emph{k}-shell attack, which will be introduced in detail in Section II-C.
\item \textbf{Rewiring:} We first release $e_{ij}$ and $e_{uv}$, and then establish two new links $e_{iv}$ and $e_{ju}$ if \textbf{Case I} is satisfied but \textbf{Case II} is not; otherwise,  establish $e_{iu}$ and $e_{jv}$ if \textbf{Case II} is satisfied but \textbf{Case I} is not. 
\end{enumerate}

Note that our rewiring will not change the degree of nodes, making the attacks \emph{imperceptible} to certain extent~\cite{schneider2011mitigation,zeng2012enhancing}.  




\subsection{Evaluation Metrics}
We use the \emph{Attack Success Rate} (ASR), \emph{Link Change Rate} (LCR), and \emph{Link Per Node} (LPN) to measure the effectiveness of various attacks. 
\begin{itemize}
    \item \textbf{ASR:} If the \emph{k}-shell of node $v_i$ is successfully changed after the attack, i.e., $k_i\neq{k^{\prime}_i}$, we set $r_{i}=1$, otherwise $r_{i}=0$, then ASR is defined as:
\begin{equation}
    \centering
    \text{ASR}=\frac{\sum_{v_i\in{V}} r_{i}}{|V|},
    \label{equ1}
\end{equation}
where $|V|$ denotes the total number of nodes in $G$. The larger the ASR is, the more successful the attack is.
\item \textbf{LCR:} If the link $e_{ij}$ is changed after the attack, we set $\psi_{ij}=1$, otherwise $\psi_{ij}=0$. then LCR is defined as:
\begin{equation}
	\centering
	\text{LCR}=\frac{\sum_{e_{ij}\in{E}}\psi_{ij}}{|E|},
	\label{equ3}
\end{equation}
where $|E|$ denotes the total number of links in $G$. LCR denotes the budget of attack, the smaller of which, the less noticeable of the attack method. 
\item \textbf{LPN:} In most cases, it would be interesting to ask that how many links do we need to change to successfully attack a node on average. We thus define LPN as follows:
\begin{equation}
	\centering
	\text{LPN}=\frac{\sum_{e_{ij}\in{E}}\psi_{ij}}{\sum_{v_i\in{V}} r_{i}}.
	\label{CEE}
\end{equation}
Similarly, the smaller of LPN, the higher effectiveness of the attack method.
\end{itemize}




\subsection{SA-based k-shell Attack}
Simulated Annealing (SA) is an optimization method, which has been used in many research areas~\cite{wu2016complex}. Our SA-based \emph{k}-shell attack is constituted of attacking iteration and annealing iteration, where the latter is nested within the former.

\textbf{Attacking:} In the attacking iteration at round $n$, we put the adversarial graph $G^{\prime}_{n-1}$ into annealing iteration (simply set $G^{\prime}_0=G$) and get $G^{\prime}_{n}$ as the output. The above attacking process will continue $N$ times, and output the final adversarial graph $G_{out}=G^{\prime}_N$.


\textbf{Annealing:} The initial and terminate temperature of the annealing are set to $T$ and $T_m$, respectively. $T_{\tau}$ represents the temperature of the system in the annealing iteration at round $\tau$, with $T_0=T$. At round $\tau$, the current temperature $T_{\tau}$ is updated to:
\begin{equation}
    T_{\tau}=\frac{T_{\tau-1}}{\tau}.
\end{equation}
We then randomly select a pair of links satisfying \textbf{Case I} or \textbf{Case II}, as introduced in Section~\ref{Sec:Def}, and rewire them on $G^{\prime}_{n-1}$ to get an adversarial graph denoted as ${G}_{a}$. 
After that, we can calculate $\text{ASR}_{\tau}$ based on ${G}_{a}$ at this round by Eq.~(\ref{equ1}). Note that, here we simply set $\text{ASR}_{0}$ as the attack success rate of $G^{\prime}_{n-1}$. 
If $\text{ASR}_{\tau}>\text{ASR}_{\tau-1}$, this rewiring operation will be accepted and recorded directly; otherwise, it will be accepted with the probability:
\begin{equation}
    p_\tau=e^{-\left|\text{ASR}_{\tau}-\text{ASR}_{\tau-1}\right|/T_{\tau}}.
\end{equation}
This process will not be terminated until $T_{\tau}\leq{T_m}$. Finally, we set $G^{\prime}_{n}={G}_{a}$ and trigger the next round of attacking.

\section{Experiments}\label{Sec:Experiment}
In this section, we compare our SA-based \emph{k}-shell attack with Random rewiring Attack (RA) and Heuristic rewiring Attack (HA), where RA refers to selecting links completely at random and HA refers to purposefully selecting two links that one is linked with nodes of high \emph{k}-shell and the other is linked with nodes of low \emph{k}-shell. We will realize the attacks based on the two-step rewiring mechanism introduced in Section~\ref{Sec:Def} and testify them on four different real-world networks.

\begin{algorithm}[!t]
\caption{\textbf{SA-based \emph{k}-shell Attack}}
\LinesNumbered
\label{alg:1}
\KwIn{the original graph $G$, the attack rounds $N$, the initial temperature $T$, the terminate temperature $T_m$.}
\KwOut{the Attacked Graph $G_{out}$.}
$n = 0$\;
$G^{\prime}_0 = G$\;
\While {$n<N$}
{
    $n++$\;
    $\tau = 0$\;
    $T_0 = T$\;
    $G^{\prime}_{n} = G^{\prime}_{n-1}$\;
    $ASR_0 = AttackSuccessRate(G^{\prime}_{n-1})$\;
    \While {$T_{\tau} > T_m$}
    {
        $Selectedlinks \gets$RandomSelectlinks\textit{$(G^{\prime}_{n-1})$}\;
    	$flag \gets$ConstraintJudge$(\textit{Selectedlinks})$\;
        \If {$flag == True$}
        {
    		$\tau ++$\;
    		${G}_{a}\gets$AttackGraph\textit{$(G^{\prime}_{n-1}, Selectedlinks)$}\;
    		
    		$ASR_{\tau} \gets$AttackSuccessRate$(\textit{${G}_{a}$})$\;
    		$T_{\tau} = \frac{T_{\tau-1}}{\tau}$\;
    		$p_{\tau}=e^{-|ASR_{\tau}-ASR_{\tau-1}|/T_{\tau}}$\;
    		\If {$ASR_{\tau} > ASR_{\tau-1}$ \textbf{or} random$(0,1) < p_{\tau}$}
    		{
    		    $G^{\prime}_{n} = {G}_{a}$\;
    		}
    		\Else
    		{
    		    $ASR_{\tau} = ASR_{\tau-1}$\;
    		}
        }
    }
}
$G_{out} = G^{\prime}_{N}$\;
\Return $G_{out}$\;
\vspace*{-1mm}
\end{algorithm}

\begin{figure*}[!t]
	\centering
	\subfigure[Karate]{
	\includegraphics[scale=0.2]{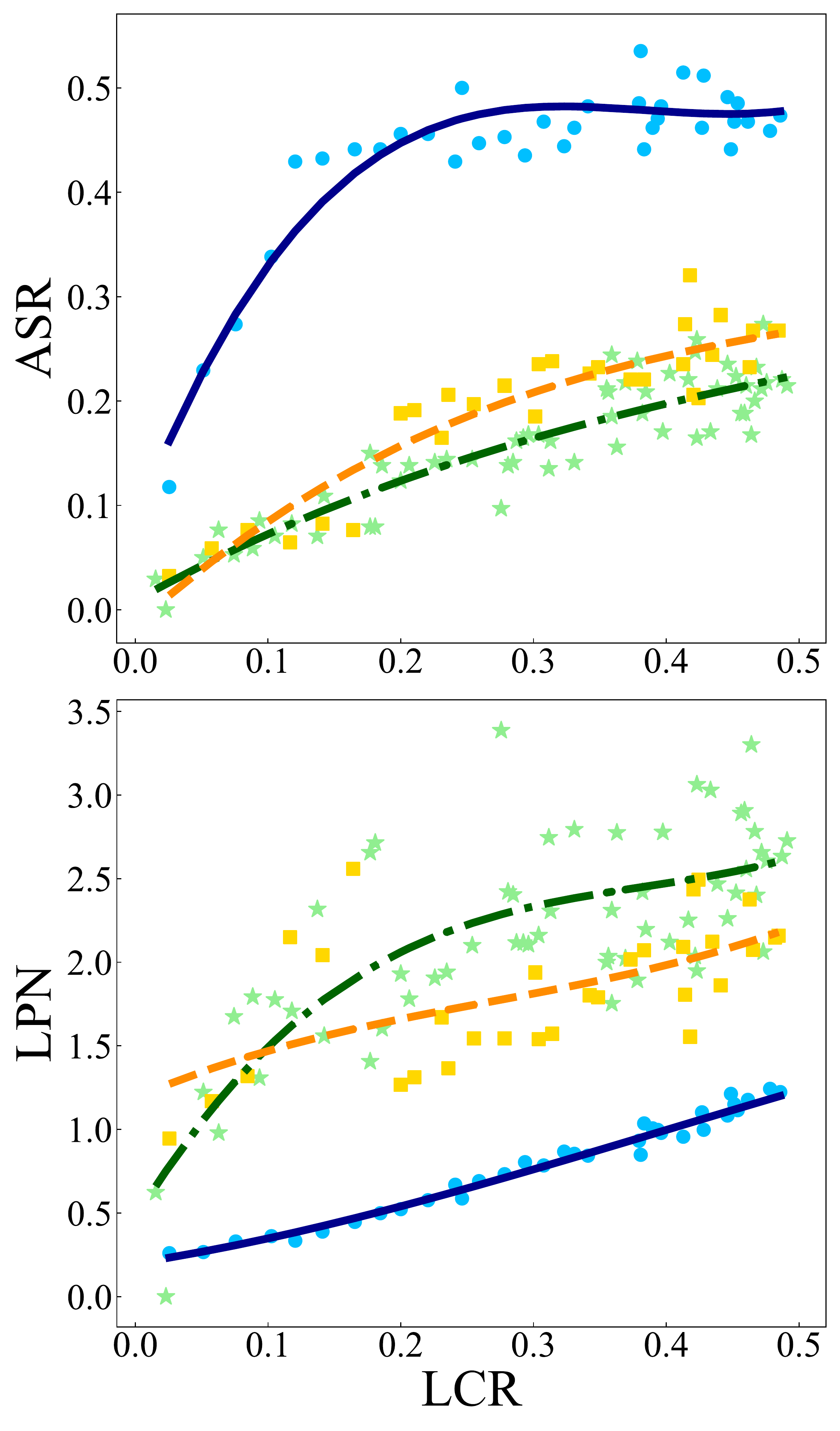}}
	\hspace{-2.5mm}
	\subfigure[Dolphin]{
	\includegraphics[scale=0.2]{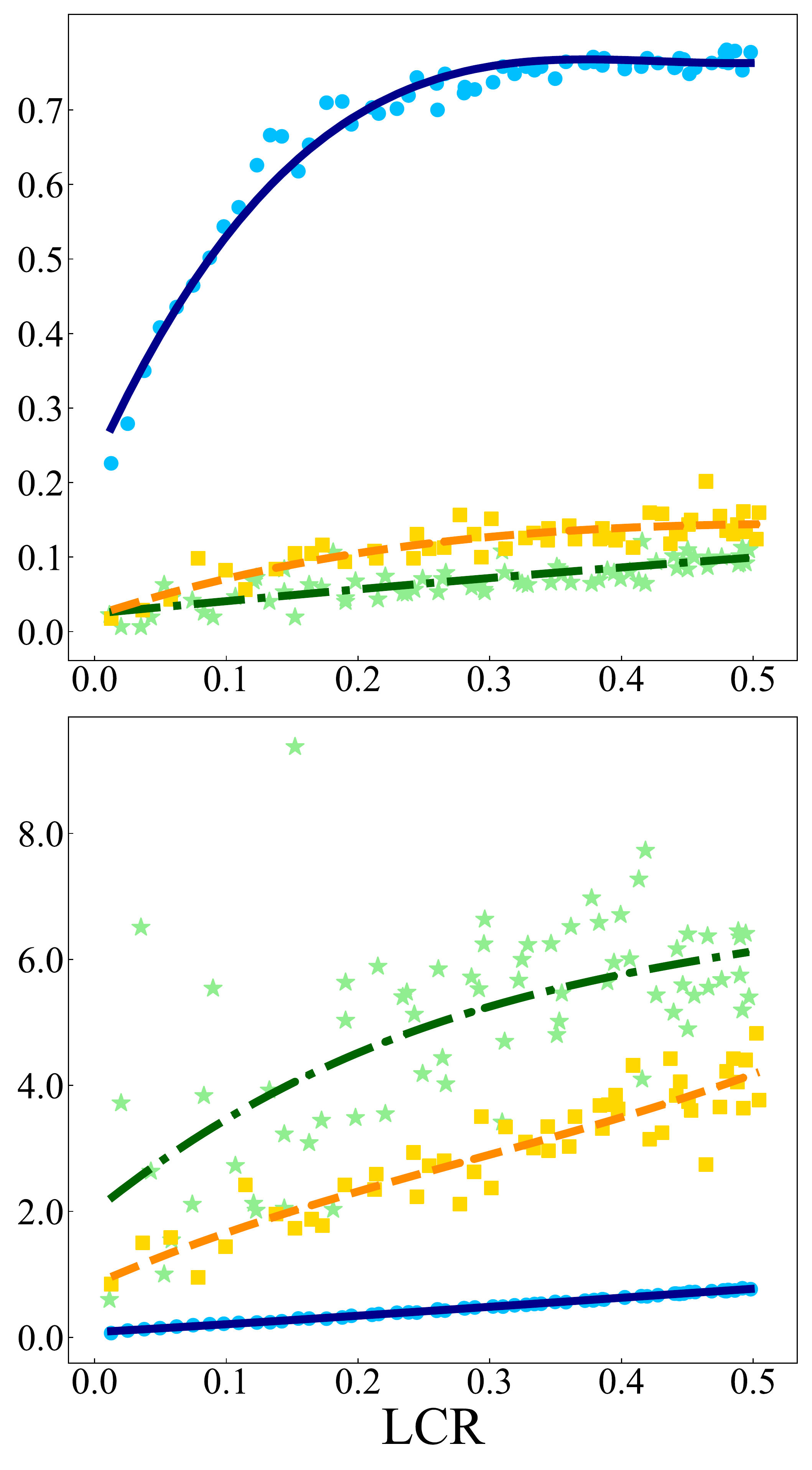}}
	\hspace{-2.5mm}
	\subfigure[Thrones]{
	\includegraphics[scale=0.2]{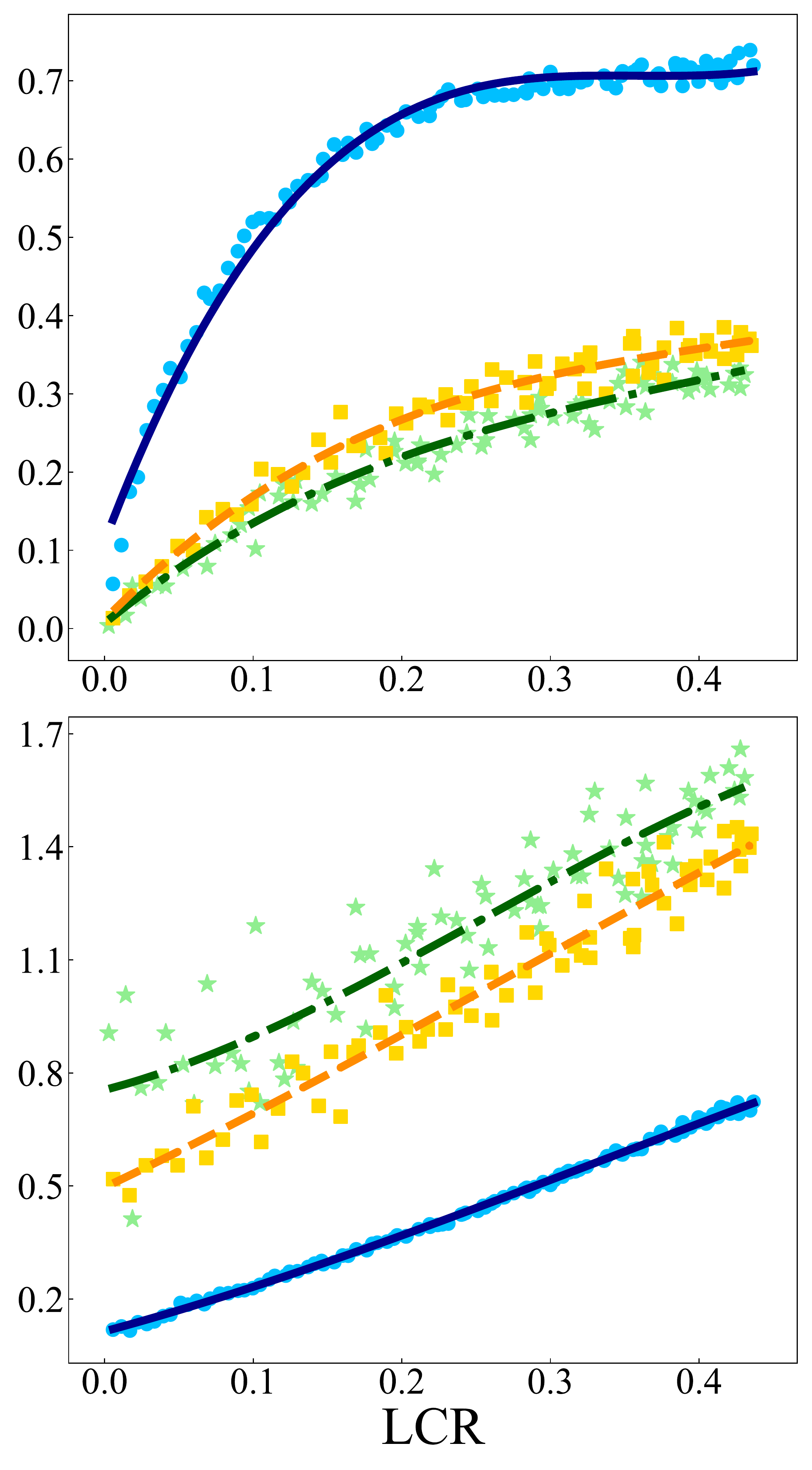}}
	\hspace{-2.5mm}
	\subfigure[Facebook]{
	\includegraphics[scale=0.2]{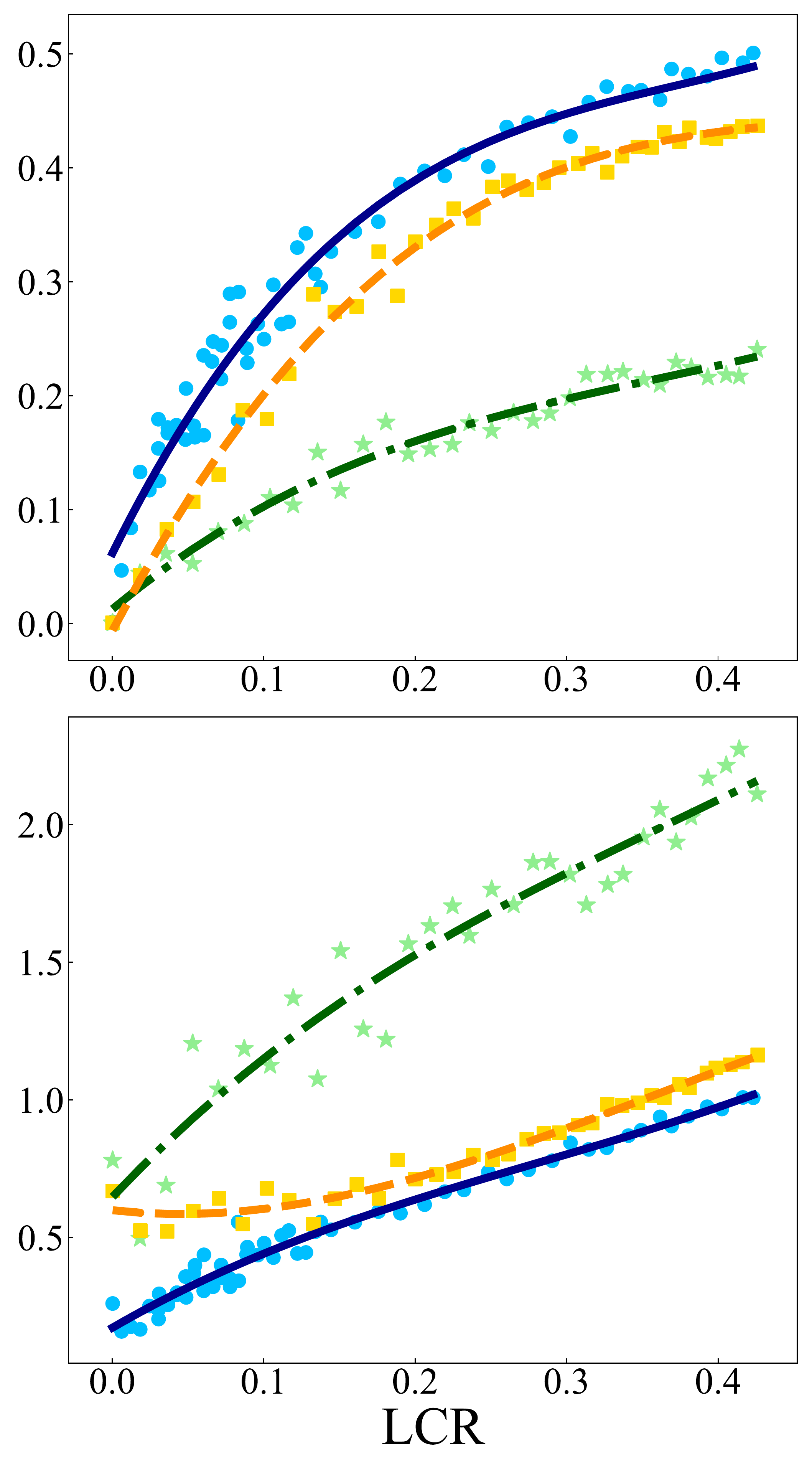}}
    \subfigure{
	\includegraphics[scale=0.25]{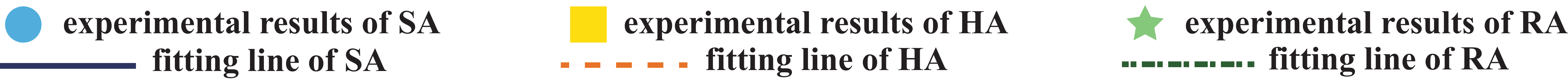}}
	\vspace{-2.2mm}
	\caption{The relationship between ASR and LCR and that between LPN and LCR on the four real-world networks by different attack methods.} 
	\label{curves}
\vspace{-5mm}
\end{figure*}
\subsection{Datasets}
The four real-world networks are briefly introduced in the following, with their basic properties presented in Table~\ref{datasets}.

\begin{itemize}
    \item \textbf{Karate:} It is a social network, which describes the interactions among the members in a university karate  recorded by Zachary~\cite{zachary1977information}.
    \item \textbf{Dolphin:} It is a social network that describes the frequent associations between bottlenose dolphins living off Doubtful Sound, New Zealand~\cite{lusseau2003bottlenose}.
    \item \textbf{Thrones:} It is a network of characters in an American fantasy drama television series, Game of Thrones~\cite{beveridge2016network}.
    \item \textbf{Facebook:} It is a online social network of students at the University of California, Irvine, which includes the users that sent or received at least one message~\cite{opsahl2009clustering}.
\end{itemize}

\vspace{-3mm}
\setlength{\tabcolsep}{3.75mm}{
\begin{table}[H]
	\centering
	\caption{Statistics of four real-world networks}
	\begin{tabular}{cccc}
		\toprule[0.5mm]
			Dataset  &  \#Nodes & \#Links & Maximum \emph{k}-shell \\ 
		\midrule
			Karate  & 34    &78     &  4 
			\\
	 	    Dolphin      & 62    & 159   &  4
	 	    \\
			Thrones      & 107   & 352   &  7
			\\
			Facebook     & 1266  & 6451  &  11
			\\
		\bottomrule[0.5mm]
	\end{tabular}
	\label{datasets}
\vspace{-3mm}
\end{table}
}

\subsection{Results and Analysis}
Next, we focus on ASR and LPN as the functions of LCR on the four networks, as shown in Fig.~\ref{curves}, based on which we can get the following results. 

Firstly, indeed, the \emph{k}-shell structure of real-world networks are much more vulnerable to adversarial attacks, compared with RA. As indicated by Kitsak et al.~\cite{kitsak2010identification}, k-shell decomposition is quite robust under random perturbation, e.g., in Dolphin network, less than 10\% nodes change their \emph{k}-shell values even when 50\% links are changed. However, it is not the case under adversarial attack, i.e., in this same network, more than 40\% nodes change their \emph{k}-shell values when only 10\% links are changed based on SA-based \emph{k}-shell attack. In general, SA-based \emph{k}-shell attack is much more successful to destroy the \emph{k}-shell structure of real-world networks than RA, while HA falls in between the two.


Secondly, SA-based \emph{k}-shell attack is also much more effective than RA and HA. To successfully change the \emph{k}-shell value of a node, only less than one link is needed to change in most cases, even when to realize the large-scale attack (corresponding to large LCR). But for RA, it may need to change more than four links to successfully attack one node in Dolphin network. Again, HA falls in between the two. Surprisingly, HA behaves quite well on attacking Facebook network, with relatively high ASR but low LPN. 
We found that, compared with the other three networks, a larger fraction of nodes in Facebook network have extremely large or small \emph{k}-shell values, i.e., 18\% of nodes have 11-shell and 19\% of nodes have 1-shell. Since HA tends to decrease those high \emph{k}-shell values but increase those low ones, it matches Facebook network quite well and thus can efficiently disturb the \emph{k}-shell values of the nodes in this network.

\begin{figure}[!t]
	\centering
	\subfigure[Original Graph]{\includegraphics[scale=0.13]{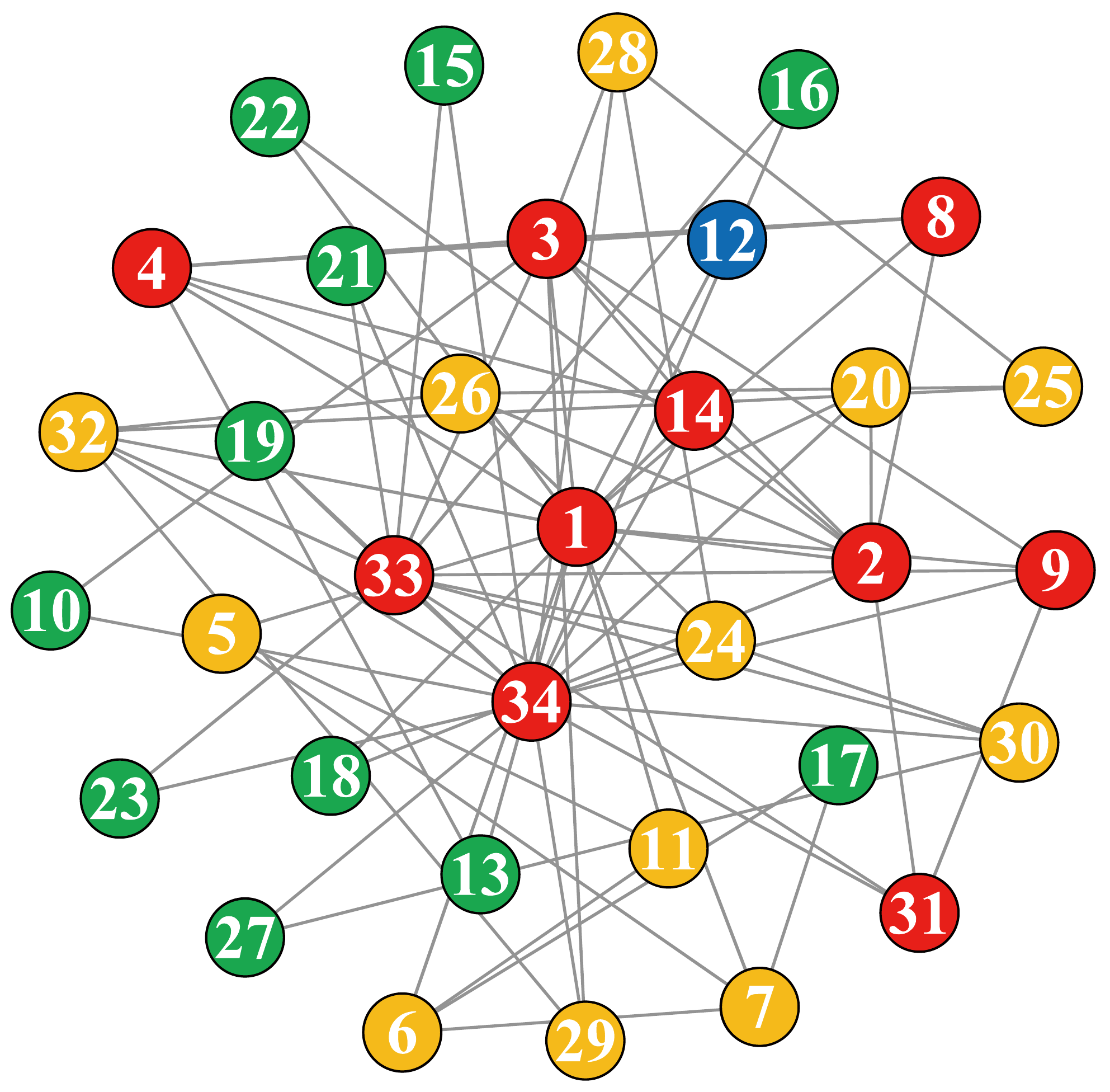}
		\label{com_graph}}
	\subfigure[$\text{LCR}=0.051$]{\includegraphics[scale=0.13]{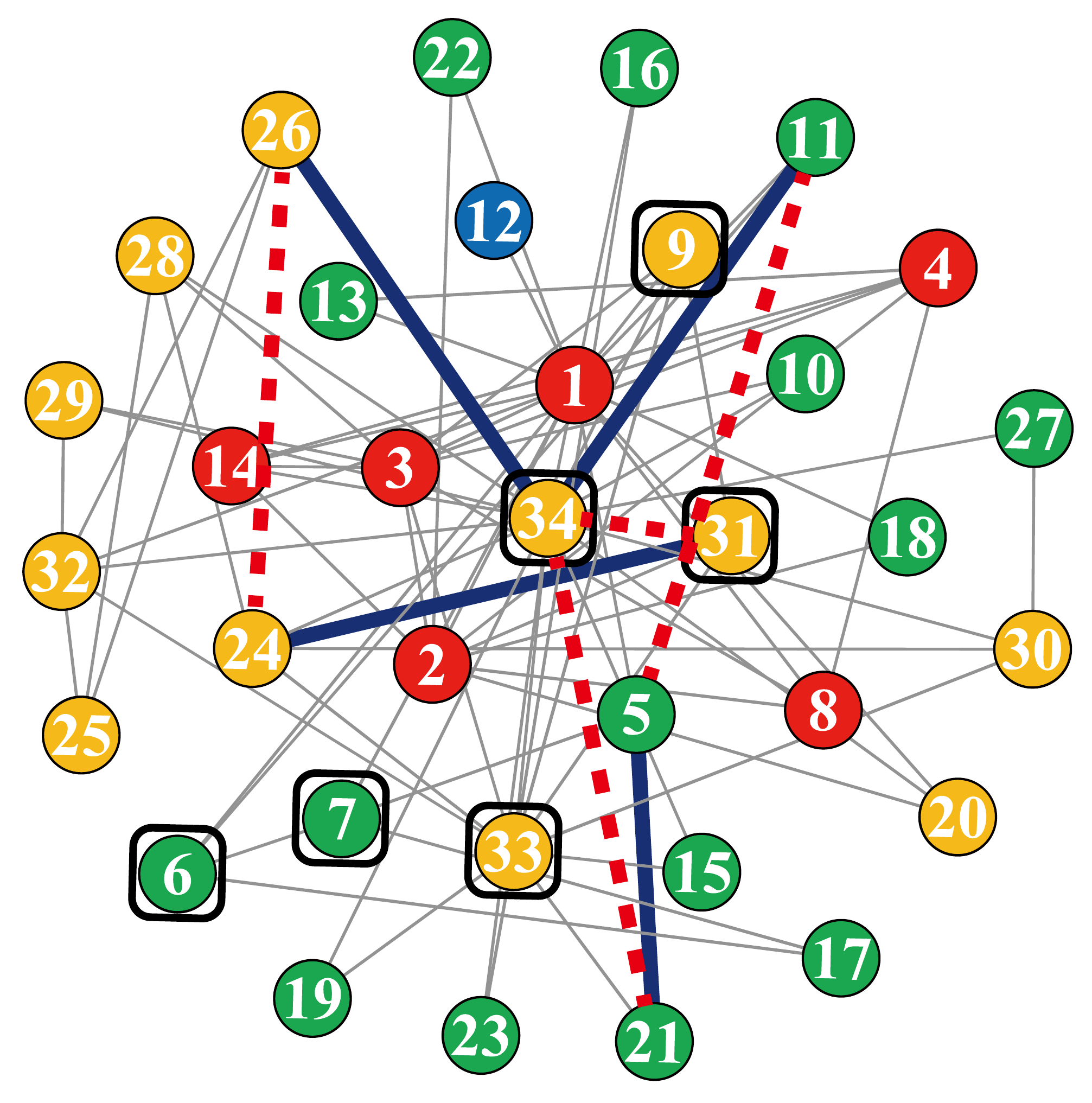}
		\label{randomdelete}}
	\subfigure[$\text{LCR}=0.205$]{\includegraphics[scale=0.13]{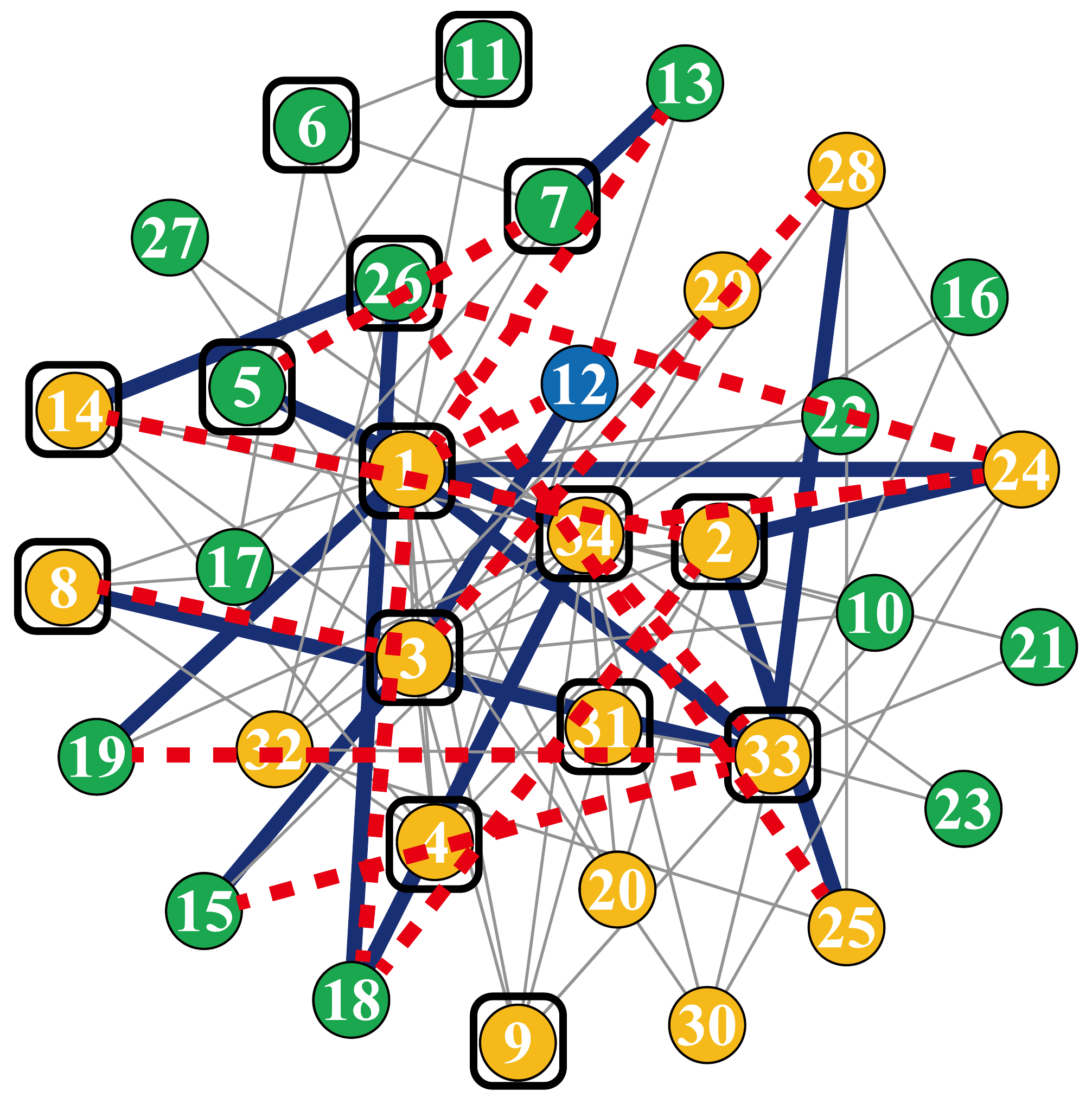}
		\label{randomreconnect}}
	\subfigure{\includegraphics[scale=0.23]{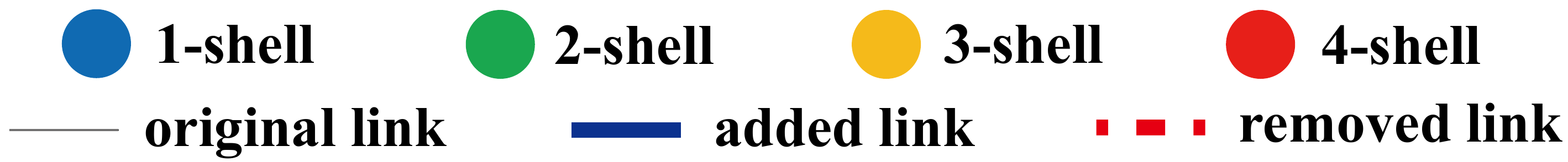}
		\label{null}}
	
	\vspace{-3mm}
	\caption{(a) The original graph. (b) and (c) represent the adversarial graphs obtained by SA-based \emph{k}-shell attack under LCR=0.051 and LCR=0.205, respectively, on Karate network.} 
	\label{figure5}
\vspace{-2mm}
\end{figure}

As a case study, we visualize the rewired links on Karate network under the SA-based \emph{k}-shell attack with different LCR, as shown in Fig.~\ref{figure5}. It can be seen that \emph{k}-shell attack is a typical combination optimization problem, i.e., different LCR values correspond to different groups of rewiring links. For instance, link $e_{26,34}$ is added to the original graph when LCR = 0.05 while it is not when LCR = 0.205. Meanwhile, we find that the \emph{k}-shell of node $v_{26}$ changes from 3 to 2 as the LCR changes from 0.051 to 0.205, while none of the links connected to it change in this process. This phenomenon suggests that we can change the \emph{k}-shell value of a target node without changing the links connected to it, making the attack quite imperceptible.

\section{conclusion}\label{Sec:Conclusion}
This paper has investigated the robustness of \emph{k}-shell structure under adversarial attacks for the first time, and has demonstrated that \emph{k}-shell decomposition is vulnerable to adversarial attacks. That is, a large fraction of nodes in a network will change their \emph{k}-shell values by only rewiring a relatively small ratio of links, using our SA-based \emph{k}-shell attack method. This phenomenon is quite different from the finding of Kitsak et al.~\cite{kitsak2010identification} that \emph{k}-shell decomposition is robust under random deletions of even up to 50\% of the links. 

Note that here we only focus on changing the \emph{k}-shell values of as many nodes as possible and the SA-based attack method is rather time-consuming. But in reality, it may be more crucial to increase or decrease the \emph{k}-shell values of a small number of target nodes more efficiently, which could be a good topic to study in the future.
\bibliographystyle{IEEEtran}
\bibliography{IEEEexample}

\begin{thebibliography}{10}
\providecommand{\url}[1]{#1}
\csname url@samestyle\endcsname
\providecommand{\newblock}{\relax}
\providecommand{\bibinfo}[2]{#2}
\providecommand{\BIBentrySTDinterwordspacing}{\spaceskip=0pt\relax}
\providecommand{\BIBentryALTinterwordstretchfactor}{4}
\providecommand{\BIBentryALTinterwordspacing}{\spaceskip=\fontdimen2\font plus
\BIBentryALTinterwordstretchfactor\fontdimen3\font minus
  \fontdimen4\font\relax}
\providecommand{\BIBforeignlanguage}[2]{{%
\expandafter\ifx\csname l@#1\endcsname\relax
\typeout{** WARNING: IEEEtran.bst: No hyphenation pattern has been}%
\typeout{** loaded for the language `#1'. Using the pattern for}%
\typeout{** the default language instead.}%
\else
\language=\csname l@#1\endcsname
\fi
#2}}
\providecommand{\BIBdecl}{\relax}
\BIBdecl

\bibitem{cook2006mining}
Q.~Xuan, Z.~Ruan, and Y.~Min, \emph{Graph Data Mining: Algorithm, Security and
  Application}.\hskip 1em plus 0.5em minus 0.4em\relax Springer, 2021.

\bibitem{1995Social}
A.~W. Wolfe, ``Social network analysis: Methods and applications,''
  \emph{Contemporary Sociology}, vol.~91, no. 435, pp. 219--220, 1995.

\bibitem{junker2011analysis}
B.~H. Junker and F.~Schreiber, \emph{Analysis of biological networks}.\hskip
  1em plus 0.5em minus 0.4em\relax John Wiley \& Sons, 2011, vol.~2.

\bibitem{2004Optimization}
G.~Paul, T.~Tanizawa, S.~Havlin, and H.~E. Stanley, ``Optimization of
  robustness of complex networks,'' \emph{The European Physical Journal B},
  vol.~38, no.~2, pp. 187--191, 2004.

\bibitem{2014Complex}
S.~Bougheas and A.~P. Kirman, \emph{Complex Financial Networks and Systemic
  Risk: A Review}.\hskip 1em plus 0.5em minus 0.4em\relax CESifo, 2014.

\bibitem{fu2018link}
C.~Fu, M.~Zhao, L.~Fan, X.~Chen, J.~Chen, Z.~Wu, Y.~Xia, and Q.~Xuan, ``Link
  weight prediction using supervised learning methods and its application to
  yelp layered network,'' \emph{IEEE Transactions on Knowledge and Data
  Engineering}, vol.~30, no.~8, pp. 1507--1518, 2018.

\bibitem{dabhi2020nodenet}
S.~Dabhi and M.~Parmar, ``Nodenet: A graph regularised neural network for node
  classification,'' \emph{arXiv preprint arXiv:2006.09022}, 2020.

\bibitem{herrgaard2008consensus}
M.~J. Herrg{\aa}rd, N.~Swainston, P.~Dobson, W.~B. Dunn, K.~Y. Arga, M.~Arvas,
  N.~Bl{\"u}thgen, S.~Borger, R.~Costenoble, M.~Heinemann \emph{et~al.}, ``A
  consensus yeast metabolic network reconstruction obtained from a community
  approach to systems biology,'' \emph{Nature biotechnology}, vol.~26, no.~10,
  pp. 1155--1160, 2008.

\bibitem{2019k}
Y.~X. Kong, G.~Y. Shi, R.~J. Wu, and Y.~C. Zhang, ``k -core: Theories and
  applications,'' \emph{Physics Reports}, vol. 832, pp. 1--32, 2019.

\bibitem{altaf2003prediction}
M.~Altaf-Ul-Amine, K.~Nishikata, T.~Korna, T.~Miyasato, Y.~Shinbo,
  M.~Arifuzzaman, C.~Wada, M.~Maeda, T.~Oshima, H.~Mori \emph{et~al.},
  ``Prediction of protein functions based on k-cores of protein-protein
  interaction networks and amino acid sequences,'' \emph{Genome Informatics},
  vol.~14, pp. 498--499, 2003.

\bibitem{2004Dynamic}
M.~Gaertler, ``Dynamic analysis of the autonomous system graph,'' \emph{in IPS
  2004, International Workshop on Inter-domain Performance and Simulation},
  2004.

\bibitem{PhysRevE.70.056122}
\BIBentryALTinterwordspacing
M.~Bogu\~n\'a, R.~Pastor-Satorras, A.~D\'{\i}az-Guilera, and A.~Arenas,
  ``Models of social networks based on social distance attachment,''
  \emph{Phys. Rev. E}, vol.~70, p. 056122, Nov 2004. [Online]. Available:
  \url{https://link.aps.org/doi/10.1103/PhysRevE.70.056122}
\BIBentrySTDinterwordspacing

\bibitem{cha2010measuring}
M.~Cha, H.~Haddadi, F.~Benevenuto, and K.~Gummadi, ``Measuring user influence
  in twitter: The million follower fallacy,'' in \emph{Proceedings of the
  international AAAI conference on web and social media}, vol.~4, no.~1, 2010.

\bibitem{2010Thresholds}
C.~Castellano and R.~Pastor-Satorras, ``Thresholds for epidemic spreading in
  networks,'' \emph{Physical Review Letters}, vol. 105, no.~21, p. 218701,
  2010.

\bibitem{2011Modeling}
G.~Bruno, P.~Nicola, V.~Alessandro, and P.~Matjaz, ``Modeling users' activity
  on twitter networks: Validation of dunbar's number,'' \emph{PLOS ONE},
  vol.~6, 2011.

\bibitem{al2017identification}
M.~A. Al-garadi, K.~D. Varathan, and S.~D. Ravana, ``Identification of
  influential spreaders in online social networks using interaction weighted
  k-core decomposition method,'' \emph{Physica A: Statistical Mechanics and its
  Applications}, vol. 468, pp. 278--288, 2017.

\bibitem{wang2021sampling}
J.~Wang, P.~Chen, B.~Ma, J.~Zhou, Z.~Ruan, G.~Chen, and Q.~Xuan, ``Sampling
  subgraph network with application to graph classification,'' \emph{arXiv
  preprint arXiv:2102.05272}, 2021.

\bibitem{Garas_2012}
\BIBentryALTinterwordspacing
A.~Garas, F.~Schweitzer, and S.~Havlin, ``Ak-shell decomposition method for
  weighted networks,'' \emph{New Journal of Physics}, vol.~14, no.~8, p.
  083030, aug 2012. [Online]. Available:
  \url{https://doi.org/10.1088/1367-2630/14/8/083030}
\BIBentrySTDinterwordspacing

\bibitem{kitsak2010identification}
M.~Kitsak, L.~K. Gallos, S.~Havlin, F.~Liljeros, L.~Muchnik, H.~E. Stanley, and
  H.~A. Makse, ``Identification of influential spreaders in complex networks,''
  \emph{Nature physics}, vol.~6, no.~11, pp. 888--893, 2010.

\bibitem{zugner2018adversarial}
D.~Z{\"u}gner, A.~Akbarnejad, and S.~G{\"u}nnemann, ``Adversarial attacks on
  neural networks for graph data,'' in \emph{Proceedings of the 24th ACM SIGKDD
  International Conference on Knowledge Discovery \& Data Mining}, 2018, pp.
  2847--2856.

\bibitem{8792200}
S.~Yu, M.~Zhao, C.~Fu, J.~Zheng, H.~Huang, X.~Shu, Q.~Xuan, and G.~Chen,
  ``Target defense against link-prediction-based attacks via evolutionary
  perturbations,'' \emph{IEEE Transactions on Knowledge and Data Engineering},
  vol.~33, no.~2, pp. 754--767, 2021.

\bibitem{chen2019ga}
J.~Chen, L.~Chen, Y.~Chen, M.~Zhao, S.~Yu, Q.~Xuan, and X.~Yang, ``Ga-based
  q-attack on community detection,'' \emph{IEEE Transactions on Computational
  Social Systems}, vol.~6, no.~3, pp. 491--503, 2019.

\bibitem{xuan2020adversarial}
Q.~Xuan, Y.~Shan, J.~Wang, Z.~Ruan, and G.~Chen, ``Adversarial attack on bc
  classification for scale-free networks,'' \emph{Chaos: An Interdisciplinary
  Journal of Nonlinear Science}, vol.~30, no.~8, p. 083102, 2020.

\bibitem{schneider2011mitigation}
C.~M. Schneider, A.~A. Moreira, J.~S. Andrade, S.~Havlin, and H.~J. Herrmann,
  ``Mitigation of malicious attacks on networks,'' \emph{Proceedings of the
  National Academy of Sciences}, vol. 108, no.~10, pp. 3838--3841, 2011.

\bibitem{zeng2012enhancing}
A.~Zeng and W.~Liu, ``Enhancing network robustness against malicious attacks,''
  \emph{Physical Review E}, vol.~85, no.~6, p. 066130, 2012.

\bibitem{wu2016complex}
J.~Wu and Y.~Xia, ``Complex-network-inspired design of traffic generation
  patterns in communication networks,'' \emph{IEEE Transactions on Circuits and
  Systems II: Express Briefs}, vol.~64, no.~5, pp. 590--594, 2016.

\bibitem{zachary1977information}
W.~W. Zachary, ``An information flow model for conflict and fission in small
  groups,'' \emph{Journal of anthropological research}, vol.~33, no.~4, pp.
  452--473, 1977.

\bibitem{lusseau2003bottlenose}
D.~Lusseau, K.~Schneider, O.~J. Boisseau, P.~Haase, E.~Slooten, and S.~M.
  Dawson, ``The bottlenose dolphin community of doubtful sound features a large
  proportion of long-lasting associations,'' \emph{Behavioral Ecology and
  Sociobiology}, vol.~54, no.~4, pp. 396--405, 2003.

\bibitem{beveridge2016network}
A.~Beveridge and J.~Shan, ``Network of thrones,'' \emph{Math Horizons},
  vol.~23, no.~4, pp. 18--22, 2016.

\bibitem{opsahl2009clustering}
T.~Opsahl and P.~Panzarasa, ``Clustering in weighted networks,'' \emph{Social
  networks}, vol.~31, no.~2, pp. 155--163, 2009.

\end{thebibliography}

\end{document}